\documentclass[
    ,final            % use final for the camera ready runs
  ]
  {aipproc}

\layoutstyle{8x11double}

\begin{document}

\title{The Hydraulic Jump in Liquid Helium}


\author{\'{E}tienne Rolley}{
  address={Laboratoire de Physique Statistique, \'{E}cole Normale Sup\'{e}rieure, 75231 
Paris, France}
}

\author{Claude Guthmann}{
  address={Laboratoire de Physique Statistique, \'{E}cole Normale Sup\'{e}rieure, 75231 
Paris, France}
}

\author{Michael S. Pettersen}{
  address={Department of Physics, Washington and Jefferson College, Washington,
PA 15301, USA}
}

\author{Christophe Chevallier}{
 address={Laboratoire de Physique Statistique, \'{E}cole Normale Sup\'{e}rieure, 75231 
Paris, France}
}

\begin{abstract}
 We present the results of some experiments on the circular hydraulic jump in
normal and superfluid
liquid helium.
 The  radius of the jump and the depth of the liquid
outside the jump are measured through
optical means.  Although the scale of the apparatus is rather small, the
	location of the jump is found to be consistent with the assumption that the
jump can be treated as a shock, if the surface tension is taken into account. 
The radius of the jump does not change when going down in temperature through
the lambda point; we think that the flow is supercritical. A
remarkable feature
of the experiment is the observation of stationary ripples within the jump when
the liquid is superfluid.  
\end{abstract}

\maketitle

%%%%%%%%%%%%%%%%%%%%%%%%%%%%%%%%%%%%%%%%%%%%
%% MAINMATTER
%%%%%%%%%%%%%%%%%%%%%%%%%%%%%%%%%%%%%%%%%%%%

%\emph{INTRODUCTION.}
When a jet of liquid falls on a horizontal surface, as one often observes in the kitchen sink, a discontinuity may occur
in the depth of the out-flowing fluid:  at a certain distance $R_j$ from the jet, there is an abrupt increase
in the depth of the liquid, and a decrease in the average velocity of the liquid.  This discontinuity is called the
hydraulic jump, and was discussed in 1914 by Rayleigh.\cite{Rayleigh}  
Although the hydraulic jump is a popular undergraduate experiment,\cite{Blackford}
the theory is challenging because of the free boundary, and it 
remains a problem of current theoretical interest.[2-7]
%The central problem associated with the phenomenon is to provide a model of $R_j$ that is both accurate and
%simple enough to reveal the physically significant parameters
%of the problem.[1-7]
%\cite{Rayleigh,Watanabe,Watson,Bohr,Godwin,Blackford,Bush}
Many experimental measurements of the
jump radius $R_j$
have been performed over the years (see the references cited in ref. \cite{Watanabe}), mostly with fluids
such as water or ethylene glycol.  
In this article, we report the results of observations of the hydraulic jump using liquid helium-4. Helium differs
from the liquids that have previously been studied in having a much lower viscosity ($\nu \sim
2 \times 10^{-8}$ m$^2$/s), so jet Reynolds numbers up to $Re \sim 4 \times 10^4$ can be achieved.  Another
difference is the scale:  typical values of $R_j$ in our apparatus are a few mm, an order or two smaller than in 
previous experiments on the hydraulic jump.  This has the effect of increasing the importance of the surface
tension.

%\emph{APPARATUS.}
The cell used in the experiment was mounted in a pumped helium optical cryostat, to permit direct imaging of the jump 
with a digital camera, as seen in Fig. 1.  
\begin{figure}
  \includegraphics[height=.2\textheight]{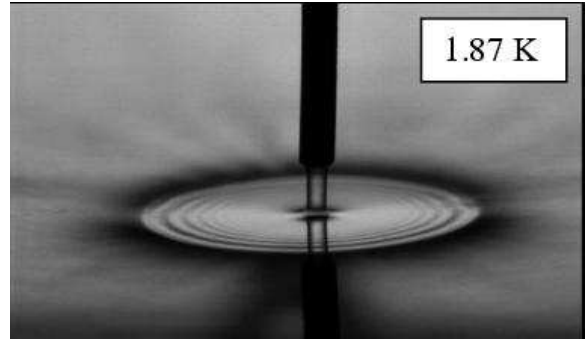}
  \caption{Image of the hydraulic jump viewed from an oblique angle.}
\end{figure}
The jet was formed by admitting helium  through a cupro-nickel capillary tube of 
100 $\mu$m inner diameter. The capillary was aligned  so that the jet would be 
perpendicular to the surface of impact.  The circular symmetry
of the jump attested to the correctness of the alignment.  The substrate surface was a
sapphire disc (chosen
for its high thermal conductivity), optically flat and aluminized.

The flux of liquid $Q$ was determined from the rate gas was admitted to the cell, measured with a flowmeter 
at room temperature.
Four heat 
exchangers (copper capillaries 0.5 m long with 1 mm inner diameter) 
cooled, liquefied and thermalized the gas.  The resulting impedance limits the maximum flow rate in the experiment.

The radius of the jump $R_j$ was determined directly from the images taken by the digital camera.  In order to permit
measurement of
the depth of the liquid, a horizontal wire was placed above the liquid.  The image of the
wire in the mirrored surface is displaced
due to the index of refraction of the liquid on the surface.   Measuring the displacement of the
image relative to the wire thus permitted a measurement
of the depth liquid outside the jump, $d$, to within a few $\mu$m.  

%\emph{RESULTS.}
Fig. 2 shows a comparison between the measurement of $R_j$  above the lambda transition and various models.
\begin{figure}
  \includegraphics[height=.25\textheight]{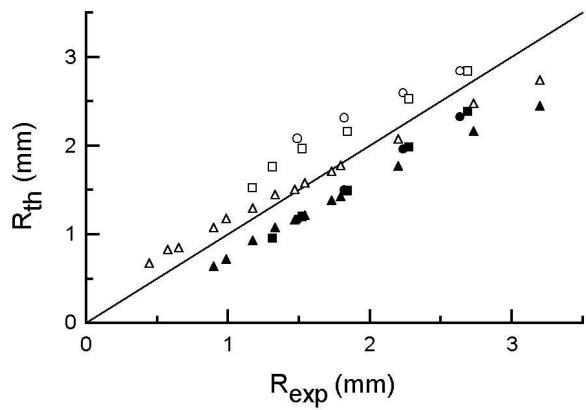}
  \caption{Experimental measurements of the hydraulic jump radius   compared with models of Watson (open symbols) and
Bush (solid symbols).  Circles:  2.45 K; squares: 3.0 K; triangles: 4.25 K.}
\end{figure}
To analyze the data, we have considered three models.  
Two of the models, those of Watson\cite{Watson} and of Bush 
and Aristoff,\cite{Bush}
follow Rayleigh in treating the jump as a shock discontinuity, where mass and momentum fluxes are conserved.
They improve on Rayleigh in including the viscosity of the liquid.  
Near the jet impact, the fluid flow is modelled as a uniform flow near the free surface, and a growing boundary
layer near the substrate; beyond the point where the boundary layer reaches the free surface, the flow is
modelled with a similarity profile.  (The matching is not exact, but close.)
Bush and Aristoff's model is similar
to Watson's, but includes the effect of the surface tension, whose curved surface exerts a force at the jump
and must be included in the momentum flux conservation condition.
The model of Bohr \emph{et al.}\cite{Bohr} differs in attempting to model the flow at the jump more accurately.
They use a one-parameter polynomial model of the flow profile; both the profile parameter and the
height of the surface are variables, which vary continuously
at the jump.  This model is capable of representing the roll which is known to
develop beyond the jump.\cite{Tani} 
%In Watson's model, 
%\begin{equation}
%{{R_j d^2 g a^2} \over {Q^2}} + {{a^2} \over {2 \pi^2 R_j d}} = 0.01676 [(R_j/a)^3 Re^{-1} +0.1826]^{-1}
%\end{equation}
%In Bush and Aristoff's model,
%\begin{displaymath}
%{{R_j d^2 g a^2}  \over {Q^2}}
%\left ( 1 + {2 \over Bo} \right )
% + {{a^2} \over {2 \pi^2 R_j d}} = 
%\end{displaymath}
%\begin{equation}
%0.01676 [(R_j/a)^2 Re^{-1} +0.1826]^{-1}
%\end{equation}
%In Bohr's model, 
%\begin{equation}
%R_j \sim 0.73 Q^{5/8} \nu^{-3/8} g^{-1/8}
%\end{equation}
%Here, $\nu$ is the kinematic viscosity, $a$ is the radius of the jet where it strikes the surface, $g$ is the acceleration of gravity,
%$Re=Q/ \nu a$ is the Reynolds number for the jet, and $Bo=\rho g R_j \Delta h / \sigma$ is the Bond number at the jump
%($\rho$ is the liquid density, $\sigma$ the surface tension, and $\Delta h$ the difference in the depth
%of the liquid across the interface).  
%The best fit to the data is provided by the model of Bush
%and Aristoff, with an agreement that is typical of  previous experiments on the hydraulic jump.

In Fig. 2 it is seen that the predictions of Watson are slightly higher than the experimental results, and
the predictions of Bush and Aristoff are slightly too low.  It would appear that the effect of
surface tension is important, but that the model of Bush and Aristoff is too crude, and overestimates the
size of the effect.  (Curiously, the experiments of Bush and Aristoff
seem to indicate that their estimate of the effect of surface tension is not large enough.)
The predictions of the model of Bohr \emph{et al.} for our experiment are similar to those of Watson;
this is not surprising in view of the fact that in practice, in all of our runs except
at the highest temperatures and lowest flow rates, the jump is rather sharp.

%\section{BELOW THE LAMBDA TRANSITION}

Below the lambda transition, one might hope to observe a transition in the jump to the value predicted by Rayleigh
in the inviscid case.
However, we did not observe such a transition.  We believe the fluid velocity in our experiment 
exceeds the critical velocity for superfluidity.  For a fluid depth on the order of 10 $\mu$m, such as we
expect within the jump,\cite{Watson} the critical velocity
is on the order of 50 mm/s,\cite{Wilks} corresponding to $Q$ less than 1 mm$^3$/s.  Unfortunately,
at such low fluxes, the jet becomes unstable and drips instead.  However, even if
the helium in our experiments is normal below the lambda point, the viscosity of the normal fluid
drops rather abruptly as the temperature is lowered from 2.17 K to 1.5 K.
In this regime, we start to observe stationary ripples inside the jump, as seen in Fig. 1.  
These
can be explained as waves 
generated at the jump and travelling upstream at the same velocity the fluid
flows downstream.  The ripple wavelength can be used to estimate the depth of the liquid inside the
jump by equating the phase velocity of shallow water capillary waves with the
fluid velocity. The resulting estimate is in rough agreement with the 
prediction\cite{Watson} and is much smaller than the depth outside the jump.  The temperature
dependence of the decay length is not yet completely understood.
Presumably,
the lower viscosity below the lambda point allows the ripples to propagate further upstream from the jump, 
so they become more prominent.
Analysis of this phenomenon is ongoing.

%\emph{CONCLUSIONS.}  
%In conclusion, 
%we have performed a measurement of the hydraulic jump in liquid helium.  Our experiment differs from previous experiments
%in the small scale of the apparatus (which increases the importance of surface tension), 
%and the low viscosity of helium (which allows us to explore a large range of Reynolds numbers).  
In summary,
our experimental results agree with the shock model of the hydraulic jump.
The effect of surface tension is important, but 
the model of Bush and Aristoff overestimates it.  No abrupt change
in the jump is observed at the lambda transition.
The appearance of ripples inside the jump below the lambda point likely reflects the decreased viscosity of the
normal fluid.

%%%%%%%%%%%%%%%%%%%%%%%%%%%%%%%%%%%%%%%%%%%%%%%%
%% BACKMATTER
%%%%%%%%%%%%%%%%%%%%%%%%%%%%%%%%%%%%%%%%%%%%%%%%

\begin{theacknowledgments}
The Laboratoire de Physique Statistique de l'\'{E}cole Normale Sup\'{e}rieure is
associated with Universit\'{e}s  Paris 6 et Paris 7.  M. S. Pettersen is grateful for the support
of the \'{E}cole Normale Sup\'{e}rieure and the Universit\'{e} de Paris 7.

\end{theacknowledgments}

%%%%%%%%%%%%%%%%%%%%%%%%%%%%%%%%%%%%%%%%%%%%%%%%
%% The bibliography can be prepared using the BibTeX program or
%% manually.
%%
%% The code below assumes that BibTeX is used.  If the bibliography is
%% produced without BibTeX comment out the following lines and see the
%% aipguide.pdf for further information.
%%
%% For your convenience a manually coded example is appended
%% after the \end{document}
%%%%%%%%%%%%%%%%%%%%%%%%%%%%%%%%%%%%%%%%%%%%%%%%

%%%%%%%%%%%%%%%%%%%%%%%%%%%%%%%%%%%%%%%%%%%%%%%%
%% You may have to change the BibTeX style below, depending on your
%% setup or preferences.
%%
%%
%% For The AIP proceedings layouts use either
%%%%%%%%%%%%%%%%%%%%%%%%%%%%%%%%%%%%%%%%%%%%

\bibliographystyle{aipproc}   % if natbib is available
%\bibliographystyle{aipprocl} % if natbib is missing

%%%%%%%%%%%%%%%%%%%%%%%%%%%%%%%%%%%%%%%%%%%
%% You probably want to use your own bibtex database here
%%%%%%%%%%%%%%%%%%%%%%%%%%%%%%%%%%%%%%%%%%%
\bibliography{sample}

%%%%%%%%%%%%%%%%%%%%%%%%%%%%%%%%%%%%%%%%%%%
%% Just a reminder that you may have to run bibtex
%% All of it up to \end{document} can be removed
%% if you don't like the warning.
%%%%%%%%%%%%%%%%%%%%%%%%%%%%%%%%%%%%%%%%%%%
%\IfFileExists{\jobname.bbl}{}
 %{\typeout{}
  %\typeout{******************************************}
%  \typeout{** Please run "bibtex \jobname" to optain}
 % \typeout{** the bibliography and then re-run LaTeX}
  %\typeout{** twice to fix the references!}
 % \typeout{******************************************}
  %\typeout{}
% }

\end{document}